\documentstyle[12pt]{article}
\begin{document}
\setlength{\baselineskip}{0.30in}
\newcommand{\be}{\begin{eqnarray}}
\newcommand{\ee}{\end{eqnarray}}
\newcommand{\bi}{\bibitem}

{\hbox to\hsize{January, 1999  \hfill TAC-1999-001}
\begin{center}
\vglue .06in
{\Large \bf {Neutrino Degeneracy and Chemically Inhomogeneous Universe}
\footnote{To be published in "Particle Physics and the Early Universe
(COSMO-98)", ed. by David O. Caldwell American Institute of Physics}}
\bigskip
\\{\bf A.D. Dolgov}
 \\[.05in]
{\it{Teoretisk Astrofysik Center\\
 Juliane Maries Vej 30, DK-2100, Copenhagen, Denmark
\footnote{Also: ITEP, Bol. Cheremushkinskaya 25, Moscow 113259, Russia.}
}}
\end{center}

\begin{abstract}

A possibility that the universe may be strongly inhomogeneous chemically,
while very smooth energetically, is considered. A possible mechanism which
could lead to such a picture is a large and inhomogeneous lepton asymmetry.
The model may explain possibly observed variation of primordial deuterium
by an order of magnitude and predicts helium-rich and possibly helium-poor
regions where the helium mass fraction differs from the observed 25\% by a
factor of 2 in either direction. Helium variation can be observed by a
different dumping rate of CMB angular spectrum at small angles at different
patches of the sky.

\end{abstract}

It is usually assumed that the universe is well described by a homogeneous
and isotropic matter distribution. Of course at the present day the matter
is quite clumpy and the mass (or energy) density contrast is large at galactic
and even at cluster scales. Averaged over larger scales the matter distribution
is probably very smooth. The observed clumpiness evolved from initially
small density fluctuations due to gravitational instability. The smoothness
of the young universe is strongly confirmed by very small angular fluctuations
of the temperature of the cosmic microwave (CMB) radiation. Though the
energetic homogeneity of the universe is well verified up to
red shift $z=10^3$, corresponding to the last scattering of CMB, its
chemical homogeneity remains an assumption, maybe quite natural, but still
an assumption. In what follows I will consider the question if a big chemical
inhomogeneity at cosmologically large scales could be compatible with the
observed smooth mass or energy distribution in the universe. We discuss a
possible mechanism which may give rise to a large variation of element
abundances in the early universe, preserving cosmological homogeneity,
$\delta \rho /\rho \ll 1$, and possible manifestations of this
variation of abundances. To a large extent my talk is based on our
paper with B. Pagel~\cite{dp}.

The starting point of this investigation from the observational side was the
reported measurements~\cite{tfb} -\cite{tbl} of the deuterium abundance
in Lyman-limit absorption line systems with red-shifts $0.48<z<3.5$ on the
line of sight to quasars. Surprisingly some groups have claimed a very
high value for the deuterium-to-hydrogen ratio, ${\rm D/H}
\approx ({\rm a\,\, few})\times 10^{-4}$. Though these results are subject
to different kinds of criticism~\cite{tbk,lev,ltb}, they put forward a very
important question: what is known about element abundances at very large
distances?

>From theoretical point of view it is an interesting challenge to find if
there exist (not too unnatural) cosmological scenarios consistent with the
observed smoothness of the universe but predicting large abundance variations.
An example of such mechanism was proposed in ref.~\cite{dk} (see
also~\cite {ad1}), where
a model of leptogenesis was considered which, first, gave a large lepton
asymmetry, which could be close to unity, and, second, this asymmetry
might strongly change on astronomically large scales, $l_L$. The magnitude of
the latter depends on the unknown parameters of the model and can easily
be in the mega-giga parsec range. The model is based on the
Affleck-Dine~\cite{afd} scenario of baryogenesis but in contrast to
the original one it gives rise
to a large (and varying) lepton asymmetry and to a small baryonic one.
There are two more models~\cite{ftv}-\cite{ccg} in the recent literature,
where a large lepton asymmetry together with a small baryonic one is
advocated, though without any significant spatial variation.

If we assume that the variation of deuterium abundance by approximately
an order of magnitude is indeed real, then the characteristic scale $l_L$
should be smaller than a gigaparsec. The lower bound on this scale
may be much smaller. It can in principle be determined by measurements
of the abundances of light elements at large distances in our neighborhood,
say, $ z\geq 0.05$.

A variation of deuterium abundance may be also explained by a variation of
the cosmic baryon-to-photon ratio. This possibility was explored in
refs.~ \cite{jf,cos}. The isocurvature
fluctuations on large scales, $l > 100$ Mpc, which are necessary to create
the observed variation of deuterium, are excluded \cite{cos} by the smallness
of angular fluctuations of the cosmic microwave background radiation (CMB).
Variations of baryonic number density on much smaller scales,
$M \sim 10^5 M_\odot$, are not in conflict with the observed smoothness of
CMB and in principle can explain the data subject to a potential conflict
with the primordial $^7$Li abundance \cite{jf}.

Exactly the same criticism of creating too large fluctuations in CMB
temperature is applicable to a simple version of the model with a varying
lepton asymmetry. One can check that the necessary value of the chemical
potential of electron neutrinos $\xi_{\nu_e}$ should be close to $-1$
to explain the possibly observed variation of deuterium by roughly an
order of magnitude. Such a change in $\xi_{\nu_e}$
with respect to zero value, assumed in our part of the world,
would induce a variation in total energy density during the RD
stage at a per cent level, which is excluded by the smoothness of CMB. However,
this objection can be avoided if there is a conspiracy between different
leptonic chemical potentials such that in different spatial regions they have
the same values but with interchange of electronic, muonic and/or tauonic
chemical potentials. Since the abundances of
light elements are much more sensitive to the magnitude of the electron
neutrino chemical potential than to those of
muon and tauon neutrinos, the variation of $\xi_{\nu_e}$
(accompanied by corresponding variations of $\xi_{\nu_\mu}$ and
$\xi_{\nu_\tau}$) would lead to a strong variation in the abundance of
deuterium and other light elements.
The equality of, say,  $\xi_{\nu_e}$ at one space point to $\xi_{\nu_\mu}$ at
another point looks like a very artificial fine-tuning, but this may
be rather naturally realized due the lepton flavor symmetry,
$e \leftrightarrow \mu \leftrightarrow  \tau$.

If lepton asymmetry changes at large distances, then not only deuterium but
also $^4 He$ would not remain constant in space. Playing with the
nucleosynthesis code~\cite{kaw} one can check that in the deuterium rich
regions the mass fraction of helium could be larger than 50\% (twice larger
than in our neighborhood). There may also exist the so called mirror regions
with a positive and large chemical potential of electronic neutrinos. In such
regions abundances of both deuterium and helium would be much smaller that
those observed nearby. For more detail see ref.~\cite{dp}. Surprisingly
nothing is known about helium abundance at large distances.  All accurate
measurements of $^4$He  based on emission lines
known to us were done at most at $z = 0.045$
corresponding  to a distance of $140 h^{-1}$ Mpc \cite{hel}, whereas helium
line and continuum absorption measurements made at high redshifts give the
abundance merely within ``a factor of a few" owing to uncertain
ionization corrections \cite{hog}. In the regions with a large fraction
of $^4 He$ one would expect bluer stars with a shorter life-time, though
the structure formation there may be inhibited due to a less efficient
cooling. In the helium poor regions the effects may be opposite. As was noted
by A. Kusenko at this Conference, a study of supernovae may help to put limits
on abundance variation. This problem definitely deserves further and more
detailed investigation.

Lepton conspiracy, mentioned above, would diminish energy density fluctuations
in first approximation. However there are some more subtle effects which
could be either dangerous for the model or observable in CMB. The first one
is related to the binding energy of $^4 He$ (7 MeV per nucleon).
Since the mass fraction
of $^4 $He may change by a factor of 2 in deuterium- (and helium-) rich regions
(from 25\% to more than 50\%),
this means that the variation in baryonic energy density may be as large as
$2\cdot 10^{-3}$. Rescaling the estimates of ref.~\cite{cos} one can
find~\cite{dp} for the fluctuations of the CMB temperature:
${\delta T / T }\approx 10^{-5} \left({R_{hor} /  10\lambda } \right)$,
where $\lambda$ is the wavelength of the fluctuation and $R_{hor}$ is the
present day horizon size. The restriction on the amplitude of temperature
fluctuations would be satisfied if $\lambda > 200 - 300 {\rm Mpc}/h_{100}$
($h_{100} = H/100$ km/sec/Mpc). Surprisingly direct astrophysical effects of
such big fluctuations of the helium mass fraction at distances above 100 Mpc
cannot be observed presently, at least the evident simple ones.

Another effect which would induce energy inhomogeneities,
is the heating of neutrinos by $e^+e^-$-annihilation at
$T\leq 1$ MeV, when neutrinos practically decoupled from plasma, and
the corresponding cooling of photons. (For the most recent and
precise calculations of this cooling see ref.~\cite{dhs}).
The efficiency of the cooling depends upon the chemical potential
of neutrinos and would create fluctuations in CMB temperature at the
level of $2\cdot 10^{-5}$~\cite{dp}.

A variation of mass fraction of primordial $^4 He$ could be observed in the
future high precision measurements of CMB anisotropies at small angular
scales \cite{hssw}. There are two possible effects, first, a slight
difference in recombination temperature which logarithmically depends on
hydrogen-to-photon ratio, and second, a strong suppression of high
multipoles with an increase of $R_p$. The latter is related to the
earlier helium recombination with respect to hydrogen and correspondingly to
a smaller number of free electrons at the moment of hydrogen recombination.
This in turn results in an increase of the mean free path of photons in
the primeval plasma and in a stronger Silk damping~\cite{js}. The position
and the magnitude of the first acoustic peak remains practically
unchanged~\cite{hssw}.

This effect seems to be very promising for obtaining a bound on or an
observation of a possible variation of primordial helium mass fraction. If
this is the case then the amplitude of high multipoles at
different directions on the sky would be quite different.
The impact of the possible variation of primordial abundances on the
angular spectrum of CMB anisotropy at low $l$ is more model dependent. It
may have a peak corresponding to the characteristic scale
$R > 200-300$ Mpc or a plateau, which would mimic the effect of the
hot dark matter.

\bigskip

{\bf Acknowledgments.}
The work of A.D. was supported by Danmarks Grundforskningsfond through its
funding of the Theoretical Astrophysical Center.

\end{document}